\documentclass[12pt]{article}
\usepackage{bbm,latexsym}
\newcommand{\be}{\begin{equation}}
\newcommand{\ee}{\end{equation}}
\newcommand{\ba}{\begin{eqnarray}}
\newcommand{\ea}{\end{eqnarray}}
\newcommand{\no}{\nonumber\\}
\newcommand{\mnu}{\mathcal{M}_\nu}
\newcommand{\deltasol}{\Delta m^2_\odot}
\newcommand{\deltaatm}{\Delta m^2_\mathrm{atm}}
\begin{document}
\title{\LARGE Zeros of the inverted neutrino mass matrix}
\author{Lu\'\i s Lavoura\thanks{E-mail: balio@cftp.ist.utl.pt} \\
\small Universidade T\'ecnica de Lisboa
and Centro de F\'\i sica Te\'orica de Part\'\i culas \\
\small Instituto Superior T\'ecnico, P--1049-001 Lisboa, Portugal}

\date{17 November 2004}

\maketitle

\begin{abstract}
I investigate viable textures with two texture zeros
for the inverted neutrino mass matrix,
and present the predictions of those textures
for the neutrino masses and for lepton mixing.
By using an Abelian symmetry
and one or two heavy scalar singlets,
I construct realizations of those textures
in the context of seesaw models.
\end{abstract}


Particle physics was highlighted in the last decade,
among other achievements,
by the discovery of neutrino oscillations and,
hence,
of the massiveness of neutrinos.
If one assumes the existence of only three light neutrinos,
then lepton mixing should be parametrized
by a $3 \times 3$ unitary matrix
\ba
U &=& \mbox{diag} \left( 1,\, e^{i \rho_1},\, e^{i \rho_2} \right)
\bar U\,
\mbox{diag} \left( e^{i \sigma_1},\,
e^{i \sigma_2},\, e^{i \sigma_3} \right),
\label{U} \\*[1mm]
\bar U &=& \left( \begin{array}{ccc}
- c_2 c_3 &
c_2 s_3 &
s_2 e^{- i \delta} \\
c_1 s_3 + s_1 s_2 c_3 e^{i \delta} &
c_1 c_3 - s_1 s_2 s_3 e^{i \delta} &
s_1 c_2 \\
s_1 s_3 - c_1 s_2 c_3 e^{i \delta} &
s_1 c_3 + c_1 s_2 s_3 e^{i \delta} &
- c_1 c_2
\end{array} \right),
\label{Ubar}
\ea
where $s_j = \sin{\theta_j}$ and $c_j = \cos{\theta_j}$
for $j = 1, 2, 3$,
the $\theta_j$ being angles of the first quadrant.
The matrix $U$  connects,
in the charged weak current $\bar \ell\, U \gamma^\mu
\left[ \left( 1 - \gamma_5 \right) / 2 \right] \nu$,
the charged-lepton fields
$\bar \ell = \left( \bar e,\, \bar \mu,\, \bar \tau \right)$
to the physical (mass-eigenstate) neutrino fields
$\nu = \left( \nu_1,\, \nu_2,\, \nu_3 \right)^T$.
In~(\ref{U}),
the phases $\rho_1$ and $\rho_2$ are unobservable,
since they can be eliminated
through rephasings of the $\mu$ and $\tau$ fields;
observable phases are only the `Dirac phase' $\delta$ and---if the
$\nu_j$ are self-conjugate fields,
as I shall assume---the `Majorana phases'
$2 \left( \sigma_1 - \sigma_3 \right)$
and $2 \left( \sigma_2 - \sigma_3 \right)$.
If one denotes the mass of $\nu_j$ by $m_j$,
then we know~\cite{tortola} that
(i) $\deltasol \equiv m_2^2 - m_1^2
\simeq 8.1 \times 10^{-5}\, \mbox{eV}^2$;
(ii) $\deltaatm \equiv \left| m_3^2 - m_1^2 \right|
\simeq 2.2 \times 10^{-3}\, \mbox{eV}^2$;
(iii) the solar mixing angle $\theta_3$ is large,
$s_3^2 \simeq 0.30$,
but far from the `maximal' value $\pi / 4$;
(iv) the atmospheric mixing angle $\theta_1$ is most likely maximal,
with $0.34 < s_1^2 < 0.68$ at $3 \sigma$ level;
(v) $\theta_2$ may well vanish,
with $s_2^2 < 0.047$ at $3 \sigma$ level.

In the weak basis where the charged-lepton mass matrix is diagonal,
the neutrino Majorana mass matrix $\mnu$,
which is symmetric,
is diagonalized by $U$ as
\be
U^T \mnu\, U = \mbox{diag} \left( m_1,\, m_2,\, m_3 \right).
\label{diagonalization}
\ee
In 2002 Frampton,
Glashow,
and Marfatia (FGM)~\cite{marfatia}
speculated that $\mnu$ may display a `texture'
such that two of its matrix elements
are zero.\footnote{Since $\mnu$ is necessarily symmetric,
$\left( \mnu \right)_{\alpha \beta}
= \left( \mnu \right)_{\beta \alpha}
= 0$ is counted as \emph{only one} zero matrix element
whenever $\alpha \neq \beta$.}
This assumption encompasses several viable possibilities;
FGM listed them,
together with the corresponding predictions for the neutrino masses
and for the parameters of the mixing matrix.
Models which embody FGM's hypothesis
have been constructed under the paradigm of $\mnu$
generated by the vacuum expectation values
of scalar triplets added to the  Standard Model
for that purpose~\cite{frigerio},
and under the paradigm of $\mnu$
generated through the seesaw mechanism~\cite{tanimoto,symmetries}.
In this mechanism,
which is much favoured on theoretical grounds,
$\mnu = - M_D^T M_R^{-1} M_D$,
where $M_D$ is the Dirac neutrino mass matrix,
which connects the flavour-eigenstate neutrinos
to some right-handed neutrinos,
and $M_R$ is the Majorana mass matrix
of those (super-heavy) right-handed neutrinos.
In the context of the seesaw mechanism,
it seems natural to assume \emph{a texture
for $\mnu^{-1}$};\footnote{I assume $\mnu$ to be non-singular.}
indeed,
zeros of $\mnu^{-1}$ are identical with zeros of $M_R$
when $M_D$ is a square diagonal matrix,
and this situation is easy to enforce in a seesaw model
with an Abelian symmetry and a relatively small number
of singlet Higgs fields~\cite{symmetries}.
It is the purpose of this letter to study
two-zero textures for $\mnu^{-1}$.

Some textures of $\mnu$ with two zeros are equivalent
to two-zero textures of $\mnu^{-1}$ \cite{tanimoto}.
This happens,
in particular,
with four textures shown to be viable by FGM:
\[
\begin{array}{rrclcrcl}
\mbox{case}\ A_1: \quad
\mnu &\sim&
{\displaystyle \left( \begin{array}{ccc}
0 & 0 & \times \\ 0 & \times & \times \\ \times & \times & \times
\end{array} \right)}
& \ \Leftrightarrow \ \
\mnu^{-1} &\sim&
{\displaystyle \left( \begin{array}{ccc}
\times & \times & \times \\ \times & \times & 0 \\ \times & 0 & 0
\end{array} \right)},
\\*[3mm]
\mbox{case}\ A_2: \quad
\mnu &\sim&
{\displaystyle \left( \begin{array}{ccc}
0 & \times & 0 \\ \times & \times & \times \\ 0 & \times & \times
\end{array} \right)}
& \ \Leftrightarrow \ \
\mnu^{-1} &\sim&
{\displaystyle \left( \begin{array}{ccc}
\times & \times & \times \\ \times & 0 & 0 \\ \times & 0 & \times
\end{array} \right)},
\\*[3mm]
\mbox{case}\ B_3: \quad
\mnu &\sim&
{\displaystyle \left( \begin{array}{ccc}
\times & 0 & \times \\ 0 & 0 & \times \\ \times & \times & \times
\end{array} \right)}
& \ \Leftrightarrow \ \
\mnu^{-1} &\sim&
{\displaystyle \left( \begin{array}{ccc}
\times & \times & 0 \\ \times & \times & \times \\ 0 & \times & 0
\end{array} \right)},
\\*[3mm]
\mbox{case}\ B_4: \quad
\mnu &\sim&
{\displaystyle \left( \begin{array}{ccc}
\times & \times & 0 \\ \times & \times & \times \\ 0 & \times & 0
\end{array} \right)}
& \ \Leftrightarrow \ \
\mnu^{-1} &\sim&
{\displaystyle \left( \begin{array}{ccc}
\times & 0 & \times \\ 0 & 0 & \times \\ \times & \times & \times
\end{array} \right)},
\end{array}
\]
where the symbol $\times$ denotes non-zero matrix elements,
and the nomenclature in the first column
is the one of FGM.
I remind that,
besides these four viable cases,
FGM found three other ones:
\[
\begin{array}{rrcl}
\mbox{case}\ B_1: \quad
\mnu &\sim&
{\displaystyle \left( \begin{array}{ccc}
\times & \times & 0 \\ \times & 0 & \times \\ 0 & \times & \times
\end{array} \right)},
\\*[3mm]
\mbox{case}\ B_2: \quad
\mnu &\sim&
{\displaystyle \left( \begin{array}{ccc}
\times & 0 & \times \\ 0 & \times & \times \\ \times & \times & 0
\end{array} \right)},
\\*[3mm]
\mbox{case}\ C: \quad
\mnu &\sim&
{\displaystyle \left( \begin{array}{ccc}
\times & \times & \times \\ \times & 0 & \times \\ \times & \times & 0
\end{array} \right)}.
\end{array}
\]
It turns out that,
besides the cases $A_{1,2}$ and $B_{3,4}$ studied by FGM,
there are three extra realistic two-zero textures for $\mnu^{-1}$:
\[
\begin{array}{rrcl}
\mbox{case}\ B_5: \quad
\mnu^{-1} &\sim&
{\displaystyle \left( \begin{array}{ccc}
\times & 0 & \times \\ 0 & \times & \times \\ \times & \times & 0
\end{array} \right)},
\\*[3mm]
\mbox{case}\ B_6: \quad
\mnu^{-1} &\sim&
{\displaystyle \left( \begin{array}{ccc}
\times & \times & 0 \\ \times & 0 & \times \\ 0 & \times & \times
\end{array} \right)},
\\*[3mm]
\mbox{case}\ D: \quad
\mnu^{-1} &\sim&
{\displaystyle \left( \begin{array}{ccc}
\times & \times & \times \\ \times & 0 & \times \\ \times & \times & 0
\end{array} \right)},
\end{array}
\]
where the nomenclature in the first column is new.

It follows from~(\ref{diagonalization}) that,
after discarding the unphysical phases $\rho_{1,2}$,
\be
\mnu = \bar U^\ast\, \mbox{diag} \left( \bar m_1,\,
\bar m_2,\, \bar m_3 \right) \bar U^\dagger,
\label{direct}
\ee
where $\bar m_j \equiv m_j e^{- 2 i \sigma_j}$.
Clearly then,
\be
\mnu^{-1} = \bar U\, \mbox{diag} \left( \frac{1}{\bar m_1}\, ,\,
\frac{1}{\bar m_2}\, ,\, \frac{1}{\bar m_3} \right) \bar U^T.
\label{inverted}
\ee
Therefore,
each two-zero texture for $\mnu$ or $\mnu^{-1}$
is equivalent to a set of two equations
\be
\begin{array}{rcl}
\bar m_1 &=& k_1 \bar m_3,
\\*[1mm]
\bar m_2 &=& k_2 \bar m_3,
\end{array}
\label{k1k2}
\ee
where $k_1$ and $k_2$ are functions of the parameters of $\bar U$.
It follows from~(\ref{k1k2}) that
\be
R_\nu \equiv \frac{\deltasol}{\deltaatm} =
\frac{\left| k_2 \right|^2 - \left| k_1 \right|^2}
{\left| 1 - \left| k_1 \right|^2 \right|}.
\ee
This quantity is experimentally known to be small,
$R_\nu \simeq 0.037$.

In order to obtain simple expressions for $k_1$ and $k_2$
it is convenient to define $\epsilon \equiv s_2 e^{i \delta}$
and to make series expansions in $\left| \epsilon \right|$,
since this parameter is experimentally known to be small.
Using the notation $t_j = \tan{\theta_j}$,
one then uses,
to first order in $\left| \epsilon \right|$,
\be
\bar U \approx \left( \begin{array}{ccc}
- c_3 &
s_3 &
\epsilon^\ast \\*[1mm]
{\displaystyle c_1 s_3 \left( 1 + \frac{\epsilon t_1}{t_3} \right)} &
{\displaystyle c_1 c_3 \left( 1 - \epsilon t_1 t_3 \right)} &
s_1 \\*[3mm]
{\displaystyle s_1 s_3 \left( 1 - \frac{\epsilon}{t_1 t_3} \right)} &
{\displaystyle s_1 c_3 \left( 1 + \frac{\epsilon t_3}{t_1} \right)} &
- c_1
\end{array} \right).
\ee
One finds that all $B$ cases yield,
to first order in $\left| \epsilon \right|$,
\be
\begin{array}{rcl}
k_1 &\approx& {\displaystyle k \left( 1 + \frac{x}{t_3} \right),}
\\*[3mm]
k_2 &\approx& {\displaystyle k \left( 1 - t_3 x \right),}
\end{array}
\label{kx}
\ee
where $k = - t_1^2$ for cases $B_1$,
$B_3$,
and $B_5$,
while
\ba
x &=& \frac{\epsilon}{t_1^3} + \frac{\epsilon^\ast}{t_1}
\quad \mbox{for\ case}\ B_1,
\\
x &=& - \frac{\epsilon}{t_1} - \epsilon^\ast t_1
\quad \mbox{for\ case}\ B_3,
\\
x &=& \epsilon t_1 + \epsilon^\ast t_1^3
\quad \mbox{for\ case}\ B_5.
\ea
The results for cases $B_2$,
$B_4$,
and $B_6$ are identical with those for cases $B_1$,
$B_3$,
and $B_5$,
respectively,
with $t_1$ substituted by $- 1 / t_1$ in both $k$ and $x$~\cite{marfatia}.
All $B$ cases yield
\be
R_\nu \approx 2 \left| k \right|^2 \left( t_3 + \frac{1}{t_3} \right)
\left| \frac{\mbox{Re}\, x}{1 - \left| k \right|^2} \right|,
\ee
and this shows that the atmospheric mixing angle
cannot be maximal in any of the $B$ cases,
since $R_\nu$ becomes too large
when $t_1$ (and hence $\left| k \right|$)
is too close to $1$.
On the other hand,
a small $\left| \epsilon \right| = s_2$
has the power to suppress $R_\nu$ in all $B$ cases.

Let us now analyze case $D$.
The relevant equations are in this case
\be
\begin{array}{rcl}
{\displaystyle \frac{\bar U_{21}^2}{\bar m_1}
+ \frac{\bar U_{22}^2}{\bar m_2}
+ \frac{\bar U_{23}^2}{\bar m_3}} &=& 0,
\\*[3mm]
{\displaystyle \frac{\bar U_{31}^2}{\bar m_1}
+ \frac{\bar U_{32}^2}{\bar m_2}
+ \frac{\bar U_{33}^2}{\bar m_3}} &=& 0.
\end{array}
\label{set}
\ee
If
\be
\begin{array}{rcl}
s_2 &=& 0,
\\
s_1 &=& c_1,
\end{array}
\label{condition}
\ee
then equations~(\ref{set}) are linearly dependent and read simply
\be
\frac{s_3^2}{\bar m_1} + \frac{c_3^2}{\bar m_2}
+ \frac{1}{\bar m_3} = 0.
\label{main}
\ee
This condition has been studied in~\cite{D4}.
It leads to a mass spectrum $m_1 < m_2 < m_3$.
The mass $m_1$ may either be of order $\sqrt{\deltasol}$
or be larger than a value of order $\sqrt{\deltaatm}$;
in particular,
an almost-degenerate mass spectrum is allowed.
If,
for definiteness,
one uses the central values $s_3^2 = 0.3$,
$m_2^2 - m_1^2 = 8.1 \times 10^{-5}\, \mbox{eV}^2$,
and $m_3^2 - m_1^2 = 2.2 \times 10^{-3}\, \mbox{eV}^2$,
then one obtains that
\be
\begin{array}{rl}
\mathrm{either} & 3.17 \times 10^{-3}\, \mathrm{eV} < m_1
< 8.28 \times 10^{-3}\, \mathrm{eV}
\\*[1mm]
\mathrm{or} & m_1 > 1.44 \times 10^{-2}\, \mathrm{eV}.
\end{array}
\label{range}
\ee
Next looking for a solution of~(\ref{set})
which does not satisfy the assumptions~(\ref{condition}),
one obtains
\be
\begin{array}{rcl}
\bar m_1 &=& {\displaystyle u\, \frac{t_3 z}{t_3 z - 1}}\, \bar m_3,
\\*[3mm]
\bar m_2 &=& {\displaystyle u\, \frac{z}{z + t_3}}\, \bar m_3,
\end{array}
\label{k}
\ee
where
\ba
u &=& \frac{- 1 + 2 \epsilon \cot{2 \theta_1} \cot{2 \theta_3}
- \epsilon^2}{c_2^2}\, ,
\label{u} \\
z &=& \epsilon \tan{2 \theta_1}.
\label{z}
\ea
Equations~(\ref{k})--(\ref{z}) are \emph{exact}.
Note that $|z|$ is not necessarily small,
since $\left| \tan{2 \theta_1} \right|$
is experimentally known to be large.
On the other hand,
and for the same reason,
$u$ is certainly very close to $-1$.
It follows from~(\ref{k}) that
\be
\frac{s_3^2}{\bar m_1} + \frac{c_3^2}{\bar m_2}
- \frac{1}{u \bar m_3} = 0,
\label{mainapprox}
\ee
an equation which is almost identical to~(\ref{main})
since $u \approx -1$.
Thus,
the approximate range~(\ref{range}) still holds.
The mixing-matrix parameter $z$ is given by
\be
z = \frac{t_3 \bar m_2}{u \bar m_3 - \bar m_2}\, .
\ee
With the $\bar m_j$ satisfying~(\ref{mainapprox}),
one obtains
\ba
\left| z \right| &=& \frac{m_1 m_2}{\sqrt{s_3^2 |u|^2 m_2^2 m_3^2
+ c_3^2 |u|^2 m_1^2 m_3^2 - m_1^2 m_2^2}}\, ,
\\
\mbox{Re}\, z &=& \frac{1}{2 c_3 s_3}\, \frac
{- s_3^4 |u|^2 m_2^2 m_3^2 + c_3^4 |u|^2 m_1^2 m_3^2
+ \left( s_3^2 - c_3^2 \right) m_1^2 m_2^2}
{s_3^2 |u|^2 m_2^2 m_3^2 + c_3^2 |u|^2 m_1^2 m_3^2 - m_1^2 m_2^2}\, .
\label{real}
\ea
The effective mass relevant for neutrinoless $\beta \beta$ decay
is in case $D$~\cite{D4}
\ba
\langle m \rangle
&=& \left| \bar m_1^\ast \bar U_{e1}^2
+ \bar m_2^\ast \bar U_{e2}^2
+ \bar m_3^\ast \bar U_{e3}^2 \right|
\no &=& \left| \frac{c_2^2 \bar m_1 \bar m_2}{u \bar m_3}
+ \bar m_3 \epsilon^2 \right|
\no &\approx& \frac{m_1 m_2}{m_3}\, .
\ea

All the textures for $\mnu^{-1}$ in this letter
can be obtained in a simple way
in models based on the seesaw mechanism;
one just needs to follow the methods of~\cite{symmetries}.
Let there be three right-handed neutrinos $\nu_{Rj}$,
which add to the standard model's
right-handed charged leptons $\ell_{Rj}$
and lepton doublets $D_{Lj} = \left( \nu_{Lj},\, \ell_{Lj} \right)^T$.
Suppose for definiteness that one wanted to produce a model
with $\mnu^{-1}$ as in case $B_5$.
One possibility
(among others~\cite{symmetries})
would consist in introducing an Abelian symmetry group $\mathbbm{Z}_4$
under which the leptons of family 1,
i.e.\ those with $j=1$,
remained invariant,
the leptons of family 2 changed sign,
and the leptons of family 3 were multiplied by $i$.
Assuming the existence of only one (the standard model's)
Higgs doublet,
and assuming that Higgs doublet to be invariant under $\mathbbm{Z}_4$,
it follows immediately from this arrangement
that both the charged-lepton mass matrix
and the neutrino Dirac mass matrix $M_D$ are diagonal.
Then,
since $\mnu^{-1} = - M_D^{-1} M_R {M_D^T}^{-1}$
in the seesaw mechanism,
zeros in $\mnu^{-1}$ are equivalent to zeros
in the right-handed-neutrino Majorana mass matrix $M_R$.
The bilinears $\nu_{Rj} \nu_{R j^\prime}$
transform under $\mathbbm{Z}_4$ as
\be
\left( \begin{array}{ccc}
1 & -1 & i \\ -1 & 1 & -i \\ i & -i & -1
\end{array} \right).
\ee
For $j = j^\prime = 1$ and $j = j^\prime = 2$
the bilinear $\nu_{Rj} \nu_{R j^\prime}$
is $\mathbbm{Z}_4$-invariant,
hence the corresponding mass terms
are directly present in the Lagrangian.
Further introducing in the theory one complex scalar singlet $S$
transforming under $\mathbbm{Z}_4$ as $S \to i S$,
we see that $S$ has an Yukawa coupling to $\nu_{R2} \nu_{R3}$
while $S^\ast$ couples to $\nu_{R1} \nu_{R3}$.
Hence,
the vacuum expectation value (VEV) of $S$
produces the $\left( 2,\, 3 \right)$ and $\left( 1,\, 3 \right)$
matrix elements of $M_R$.
The matrix elements $\left( M_R \right)_{33}$
and $\left( M_R \right)_{12}$ remain zero,
as required by the texture $B_5$,
since they violate $\mathbbm{Z}_4$
and since there is no scalar singlet
with Yukawa couplings appropriate to generate them.

Case $B_6$ can be produced in an analogous way to case $B_5$.
For case $D$,
one needs once again a symmetry $\mathbbm{Z}_4$
and a complex singlet $S$ transforming into $i S$
under $\mathbbm{Z}_4$.
The first-generation leptons are $\mathbbm{Z}_4$-invariant,
the second-generation ones transform with $i$,
and the third-generation ones with $-i$.
The bilinears $\nu_{Rj} \nu_{R j^\prime}$
now transform as
\be
\left( \begin{array}{ccc}
1 & i & -i \\ i & -1 & 1 \\ -i & 1 & -1
\end{array} \right),
\ee
and therefore the $\left( 2,\, 2 \right)$ and $\left( 3,\, 3 \right)$
matrix elements of $M_R$ are zero.

Cases $A_{1,2}$ and $B_{3,4}$ are more demanding;
I dwell on case $A_1$ for definiteness.
I introduce a symmetry $\mathbbm{Z}_8$
under which the leptons of the first family remain invariant,
those of the second family change sign,
and those of the third family are multiplied
by $\omega = \exp{\left( i \pi / 4 \right)}$.
The Higgs doublet is,
as before,
unique and invariant under this $\mathbbm{Z}_8$ symmetry.
The $\nu_{Rj} \nu_{R j^\prime}$ transform as
\be
\left( \begin{array}{ccc}
1 & \omega^4 & \omega \\ \omega^4 & 1 & \omega^5 \\
\omega & \omega^5 & \omega^2
\end{array} \right).
\ee
The $\left( 1,\, 1 \right)$ and $\left( 2,\, 2 \right)$
matrix elements of $M_R$ are $\mathbbm{Z}_8$-invariant.
The $\left( 1,\, 2 \right)$ matrix element
requires the presence of a real scalar singlet
which changes sign under $\mathbbm{Z}_8$.
The $\left( 1,\, 3 \right)$ matrix element
is generated by the Yukawa coupling to a complex scalar singlet
which gets multiplied by $\omega^7$ under $\mathbbm{Z}_8$.
The other entries of $M_R$ remain zero
in the absence of any further scalar singlets.

Thus,
all the textures for $\mnu^{-1}$ advocated in this letter
can be easily justified by seesaw models
supplemented with an Abelian symmetry
and one or two scalar singlets with VEVs at the seesaw scale.
This implementation of the textures
operates both at the seesaw scale and at any other scale.
Indeed,
in the standard model \emph{with only one Higgs doublet},
the matrices $\mnu$ at any two energy scales $\mu_1$ and $\mu_2$
are related by~\cite{chankowski}
\be
\mnu \left( \mu_1 \right) = I \mnu \left( \mu_2 \right) I,
\label{relation}
\ee
where the matrix $I$
(which depends on $\mu_1$ and $\mu_2$)
is \emph{diagonal},
positive,
and non-singular.
It follows from~(\ref{relation}) that
any zero matrix element of $\mnu$,
or of $\mnu^{-1}$,
at a given energy scale,
remains zero at any other energy scale.

To summarize,
I have shown in this letter that two-zero textures
for the inverted neutrino mass matrix
are quite easy to obtain in the context of seesaw models.
There are seven such textures which are not in disagreement
with the available data on neutrino masses and lepton mixing;
four of those seven viable textures
coincide with textures for $\mnu$ previously studied by FGM.
The other three textures are new,
and one of them in particular---called `case $D$'
in this letter---produces the interesting prediction~(\ref{main})
for the neutrino masses and Majorana phases.


\vskip 5mm
\noindent \textbf{Acknowledgements}:
I thank Walter Grimus for reading the manuscript.
This work was supported by the Portuguese
\textit{Fun\-da\-\c c\~ao para a Ci\^encia e a Tecnologia}
under the project U777--Plurianual.


\end{document}